\documentclass[twocolumn,showpacs,preprintnumbers,amsmath,amssymb]{revtex4}
\usepackage{graphicx}
\usepackage{dcolumn}
\usepackage{bm}

\begin{document}

\title{Thermal Model Description of Strangeness Enhancement\\
at Mid-Rapidity in\\
Pb--Pb collisions at 158 GeV A/c.\\}
\author{Sahal Yacoob}%
\altaffiliation[ Also at ]{Department of Physics and Astronomy, Northwestern University}
 \email{sahal@northwestern.edu}
\author{Jean Cleymans}
 \email{Cleymans@physci.uct.ac.za}
\affiliation{%
Physics Department \\
University of Cape Town \\
Private Bag 7700 \\
Rondebosch \\
South Africa \\
}
\date{August 26,2002}
\begin{abstract}
The results of the WA97 collaboration for strange particle
production at mid-rapidity in Pb--Pb collisions at 158
GeV$\cdot$A/c at CERN  display a strong strangeness enhancement
with system size at mid-rapidity which is dependent on the
strangeness of the particle concerned, and saturates at values of
participating nucleons greater than 120. These results are
phenomenologically described by the mixed canonical ensemble, with
canonical (exact) strangeness conservation involving all strange
resonances, and grand canonical conservation of charge and baryon
number. A detailed quantitative analysis shows that the data are
well described by an equilibrium $(\gamma_S \equiv 1) $ hadron
gas.

\end{abstract}

\maketitle

\section{\label{sec:Intro}Introduction\protect\\}
The WA97 collaboration\cite{59} at CERN
has shown that strange particle yields per wounded
nucleon reach a saturation level for the most
central Pb-Pb collisions (the Pb-Pb data
is presented in 4 centrality bins)
and show a pronounced increase when compared to a p-Be system.
A mixed canonical description of these data has been
published by Hamieh et al.\cite{hamieh,stonybrook} who showed that the density
of strange particles derived from such a treatment depends on the
size of the system consistent with the WA97 observations\cite{59}.
In addition Hamieh et al. \cite{hamieh}
have shown that if reasonable values are used
for the thermal parameters of each system the behavior of the
yield (in the Pb-Pb systems) per wounded nucleon normalized to
the p-Be yield is in agreement with the values obtained by the
WA97 collaboration. The present work uses
also he mixed canonical
ensemble \cite{Braun-Munzinger}, and
presents a more thorough analysis
in that the thermal parameters for each system are obtained
by minimizing $\chi^2$ and a more
explicit comparison between model and data is performed.
The dependence of the parameters on the size of the system is
presented, as well as the yields per wounded nucleon relative to p-Be.
\section{Mixed Canonical Formalism}
The mixed canonical partition function for a hadron gas
conserving strangeness exactly is described by the following partition
function
\cite{formal}:
\begin{equation}
Z^C_{S=0}=\frac{1}{2\pi}
       \int_{-\pi}^{\pi}
    d\phi~ \exp{\left(\sum_{n=- 3}^3S_ne^{in\phi}\right)},
\end{equation}
where $S_n= V\sum_k Z_k^1$, $V$ is the volume, and the sum
is over all particles and resonances carrying strangeness $n$.
For a particle of  mass $m_k$, with spin-isospin degeneracy factor
$g_k$, carrying  baryon number $B_k$ and charge $Q_k$ with baryon chemical potential $\mu_B$ and charge chemical potential $\mu_Q$,
 the one-particle partition function is expressed in the Boltzmann
approximation as:
\begin{equation}
 Z_k^1\equiv {\frac{g_k}{2\pi^2}}
m_k^2TK_2\left({\frac{m_k}{T}}\right)\exp (B_k\mu_B+Q_k\mu_Q) . \label{e:zk1}
\end{equation}
As described in Ref. \cite{Braun-Munzinger}
the partition function for this ensemble may be rewritten
in the following form which is well suited for numerical calculation:
\begin{eqnarray}
Z^C_{S=0}=e^{S_0}
\sum_{n=-\infty}^{\infty}\sum_{p=-\infty}^{\infty} a_{3}^{p}
a_{2}^{n} a_{1}^{{-2n-3p}} \nonumber \\
\times I_n(x_2) I_p(x_3) I_{-2n-3p}(x_1)
 , \label{eq2}
\end{eqnarray}
where
\begin{eqnarray}
a_i&=& \sqrt{{S_i}/{S_{-i}}}\\
x_i &=& 2\sqrt{S_iS_{-i}}
\end{eqnarray}
and $I_i$ are modified Bessel functions.

The expression for the  particle density, $n_i$, may be obtained
from the partition function Eq.~(3) by following the standard
method \cite{formal}. For a particle $i$ having strangeness $s$ the
result is
\begin{eqnarray}
n_{i}={\frac{Z^1_{i}}{Z_{S=0}^C}}
\sum_{n=-\infty}^{\infty}\sum_{p=-\infty}^{\infty} a_{3}^{p}
a_{2}^{n}
 a_{1}^{{-2n-3p- s}} \nonumber \\
\times I_n(x_2) I_p(x_3) I_{-2n-3p- s}(x_1)
 . \label{eq5}
\end{eqnarray}
\section{Results}
In this section we explore whether canonical strangeness suppression
at small volumes (compared to grand canonical equilibrium particle yields)
is able to explain the enhancement of strange particle
yields per wounded nucleon from small to large systems, as measured
by the  WA97 collaboration at CERN\cite{59}.
The thermal model is well suited for to
$4\pi$ integrated data  and to the central rapidity region
if a boost invariant
plateau exists around at mid-rapidity\cite{cleymans-redlich}.
We assume the validity of
the latter in order
to analyze the  particle yields.
A thorough comparison of $4\pi$  particle numbers
with the thermal model has been made
 in Ref. \cite{NA49}.
As has been the case in other calculations \cite{stachel,stachel2}, our
findings at mid-rapidity are found to
be consistent with full  strangeness equilibration (i.e. $\gamma_s=1$).
This is in contrast to $4\pi$ integrated particle yields which
are incompatible with
full strangeness chemical equilibriun and deviate from it by several standard
deviations.
Experimentally, ratios of particle to
anti-particle yields\footnote{The reason for checking particle
to anti-particle ratios is that the masses of the particles in the
ratio are equal -  this leads to these particles being equally affected
by flow. This in turn  minimizes the errors introduced by
considering a limited kinematic region \cite{gorenstein2}}
at mid-rapidity\footnote{In symmetric collisions the majority of new
particles are produced at mid-rapidity.} have been compared to
$4\pi$ integrated yields and shown to be in agreement for S+S
collisions at $200$ A GeV by the CERN NA35 collaboration \cite{S+S}.
In addition, and perhaps of greater interest, the $4\pi$ integrated
ratio $\overline{\Xi}/\Xi$ measured by CERN NA49 for Pb+Pb collisions
at 158 A GeV has been shown to agree with the corresponding
CERN WA97 mid-rapidity ratio \cite{barton}.

As the mixed canonical formalism has been derived here for the case
of Boltzmann statistics, it is worth noting that the second term in
a series expansion of the correct quantum statistical distribution
function for kaons gives a correction to the kaon numbers of the order
of $3 \%$ at a temperature of 150 MeV.
The corrections to the Boltzmann distribution
functions due to quantum mechanics are expected to affect the kaons
more than any other strange particle, as they are the lightest of
the strange particles.
With errors of $3 \%$  and less, the use of Boltzmann statistics to
describe the strange particles is justified.
For all mixed canonical analyses the parameter $\mu_Q$ fit to zero, and
has  been subsequently removed as a free parameter. For a $4 \pi$
integrated $\mathrm{Pb}+\mathrm{Pb}$ system, $\mu_Q$ is expected to be
small and negative.

\subsection{Analysis A}
The main results of analysis A are based on fitting the ratios of
the strange particle yields to the yield of negatives for
each case in Table \ref{table:results}.
The negatives yield is used for each ratio due to the
good statistics.
In addition it provides a measure of the entropy of the collision.
For negative pions, the
 quantum mechanical correction to the Boltzmann distribution function at
150 MeV is of the order of $30\%$ .
It is unacceptable to ignore this correction
when attempting to describe the data. Considering that
the non-strange particle yields predicted by the model are independent
of the exact conservation of strangeness, the full quantum mechanical
grand canonical particle number expression is evaluated for all
non-strange particles.
The parameters obtained in this way are shown in Table \ref{table:results}
as analysis A\footnote{The table also includes the $\chi^2$ and fit
parameters for two other analyses described in detail below}.
The $\chi^2$ for each system is good, and there are three (two)
degrees of freedom for the p+Pb, and Pb+Pb systems (p+Be system).
The large uncertainties in the radius parameter for the Pb+Pb systems
shows that we have hit the grand canonical limit, and have no volume
dependence, this is discussed briefly in a later section.
\subsection{Analysis B}
Analysis B has only three data
points ($ \frac{\overline{\Lambda}}{\Lambda},\frac{\overline{\Xi}}{\Xi} $
and $\frac{\overline{\Xi}}{\overline{\Lambda}}$)  for each fit.
This analysis has been motivated by the reasons mentioned
previously in the text.
This analysis has no
free parameters and shows reasonable agreement with analysis A.
The large uncertainties of the fitted parameters are unavoidable.
Interestingly, fitting these ratios proves impossible for the
p+Pb system. The hypothesis of \cite{hamieh} of a special
`interaction volume' for strange particles in this system cannot
be tested, as the particle ratios would depend only on this
interaction volume, as the regular volume cancels in the ratio of
the particle multiplicities.

\subsection{Analysis C}
Analysis C determines the fits to the data
using the grand canonical ensemble with
quantum statistics. In this case, the volume dependence
canels out in the
particle ratios and we have included $\mu_S$ and $\mu_Q$. This procedure
fails for the p+Be system. As may be expected for a system of this size, a
canonical treatment \cite{marais} is required. In this analysis, the
parameters fitted for Bin 4 immediately catch the eye -- with high
uncertainty values even though $\chi^2 < 1$. This is especially puzzling
when one notices that this is not the case in analyses A or B. The $T$
and $\mu_B$ parameters agree with those obtained by
Becattini et al.~\cite{NA49}, for a grand canonical description of
the $4\pi$-integrated data for central symmetric $\mathrm{Pb}^{208}$
collisions at 158 A GeV by the CERN NA49 collaboration.
\begin{table*}
\caption{\label{table:results} Parameters obtained for various
analyses as described in the text. The temperature ($T$) and chemical
potential ($\mu_B$) are in MeV. The radii ($R$) are given in fm.
If $\chi^2$ is greater than 50, the fit parameters are not shown.}
\begin{tabular}{|c|r@{$\pm$}l|r@{$\pm$}l|r@{$\pm$}l|r@{$\pm$}l|r@{$\pm$}l|r@{$\pm$}l|}
\hline
&\multicolumn{2}{c|}{p+Be} &
\multicolumn{2}{c|}{p+Pb} &
\multicolumn{8}{c|}{Pb+Pb} \\
\cline{6-13}
&\multicolumn{2}{c|}{} &
\multicolumn{2}{c|}{} &
\multicolumn{2}{c|}{Bin1}&
\multicolumn{2}{c|}{Bin2}&
\multicolumn{2}{c|}{Bin3}&
\multicolumn{2}{c|}{Bin4}\\
\hline
\multicolumn{13}{|c|}{Analysis A}\\
\multicolumn{13}{|c|}{Ratios of Strange Particle to Negatives Multiplicities}\\
\hline
$\chi^2$ & \multicolumn{2}{c|} {0.756}& \multicolumn{2}{c|} {1.85 }& \multicolumn{2}{c|} {2.07}& \multicolumn{2}{c|} {0.427}& \multicolumn{2}{c|} {1.58}& \multicolumn{2}{c|} {1.45}\\
$T$ &162&4& 172 & 2 & 161 & 4 & 164 & 4 & 166 & 4  & 159  & 5  \\
$\mu_B$ &111&7& 157 & 39 & 201 & 25 & 230 & 38& 228 & 28 & 204 & 30 \\
$R$ & 1.39 & 0.14 &  1.16  & 0.39 & 6.81 & 8.73 & 9.88 & 10.9 & 6.78 & 9.39 & 10.4 & 6.8 \\
\hline
\multicolumn{13}{|c|}{Analysis B}\\
\multicolumn{13}{|c|}{Ratios of Particle Yields to that of their Anti-Particles}\\
\hline
$\chi^2$ & \multicolumn{2}{c|} {$ \sim 10^{-6}$ }& \multicolumn{2}{c|} {60287 }  & \multicolumn{2}{c|} { $\sim 10^{-7}$}& \multicolumn{2}{c|} { $\sim 10^{-4}$}& \multicolumn{2}{c|} {$ \sim 10^{-3}$}& \multicolumn{2}{c|} {$ \sim 10^{-2}$}\\
$T$ &157    &25  & \multicolumn{2}{c|}{-} & 171 & 17 & 161 & 15 & 153 & 35 & 163  & 45  \\
$\mu_B$ & 106  & 31&   \multicolumn{2}{c|}{-} & 229 & 55 & 222 & 46& 190 & 61 & 220 & 30 \\
$R$ & 1.50 & 0.83 &   \multicolumn{2}{c|}{-} & 2.17 & 2.03 & 6.06 & 8.92& 9.38 & 7.12 & 8.08 & 6.83 \\
\hline
\multicolumn{13}{|c|}{Analysis C}\\
\multicolumn{13}{|c|}{Grand Canonical Fit}\\
\hline
$\chi^2$ & \multicolumn{2}{c|} {78}& \multicolumn{2}{c|} {17.5 }& \multicolumn{2}{c|} {1.53}& \multicolumn{2}{c|} {0.15}& \multicolumn{2}{c|} {0.70}& \multicolumn{2}{c|} {0.97}\\
$T$ &\multicolumn{2}{c|}{-} & 143 & 8 & 173 & 20 & 165 & 4  & 166  & 5 & 171&20\\
$\mu_B$ &  \multicolumn{2}{c|} {-} &205 & 106& 282 & 96 & 227 & 39& 203 & 45 & 265 & 120\\
$\mu_{Q}$ &\multicolumn{2}{c|}{-} & -109 & 52 & -54 & 68& 0 & 50 & 0  & 60& -62& 86\\
$\mu_S$ &  \multicolumn{2}{c|} {-} & 105& 63 & 110 & 73 & 54 & 25& 39 & 28 & 94 & 97 \\
\hline
\end{tabular}
\end{table*}
The parameters in analyses A, B, and C agree within th each other one standard
deviation.
\subsection{Thermal Model Parameters}
In order to fully reproduce the data obtained by
the WA97 collaboration , it is necessary to
determine a relationship between the
parameters $\mu_B$, the radius $R$ and the temperature  $T$ obtained
 from the fits of
analysis A and the number of wounded nucleons.
The fits used are shown in Figs. \ref{fig:temp} (temperature),
and \ref{fig:mub} ($\mu_B$),  as well as the parameters obtained
from a full Boltzmann treatment\footnote{Using Boltzmann statistics
for strange, and non-strange particles}. The volume paramaterisation
is described by equation \ref{eq:radius}.
The parameters obtained from the p+Pb data have not been considered
because of their prediction of a volume smaller than that of the
p+Be system.
\subsubsection{Temperature}
\begin{figure}
\includegraphics[width=\linewidth]{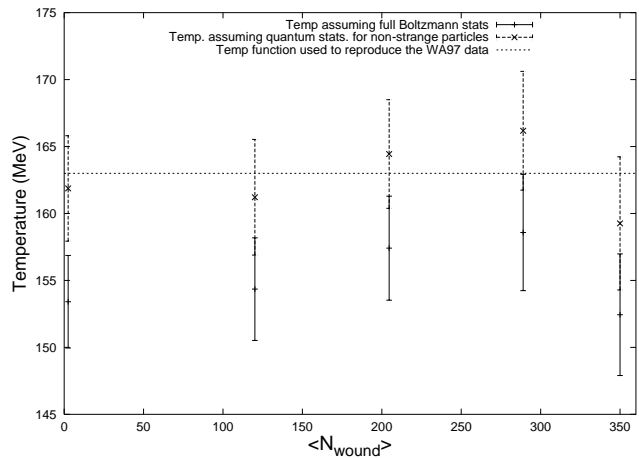}
\caption{\label{fig:temp} Variation of the chemical
freeze-out temperature ($T$) of the system with number of wounded nucleons.}
\end{figure}

As may be seen in Fig. \ref{fig:temp}, it appears initially that
the temperature increases with centrality. However, the last data point
does not fit this hypothesis, and the variation of chemical freeze-out
temperature with number of participants has been assumed to be
constant ($T = 163$ MeV). This is in agreement with the
fit to the NA49 data by Becattini et al. \cite{NA49} to central
Pb+Pb data.
This temperature value is slightly less than the one obtained
by Hamieh et al. \cite{hamieh}.

\subsubsection{Radius}
In order to reproduce the
WA97 data, knowledge of the variation of system size with average
number of wounded nucleons is required. We have chosen a function of the form:
\begin{equation}
<N_{\mathrm{wound}}> = R^3 + b
\label{eq:radius}
\end{equation}
with  $b<0.5$ adjusted to reproduce the data of the WA97 collaboration
(Figs.~\ref{fig:WA97fa} and \ref{fig:WA97f}). This differs from the
dependence of the radius on $A_{part}$ assumed by
Hamieh et al. \cite{hamieh} where $A_{part} \sim 1.3 - 1.7 R^3$.

The large uncertainty in the radius for the larger systems is not
surprising, as the ratios being fitted are expected to show a small volume
dependence, due only to canonical strangeness suppression. The effects
of canonical strangeness suppression decrease with system size as one
approaches the grand canonical limit where ratios of particle multiplicities
have no volume dependence.

\subsubsection{Baryon Chemical Potential}
\begin{figure}
\includegraphics[width=\linewidth]{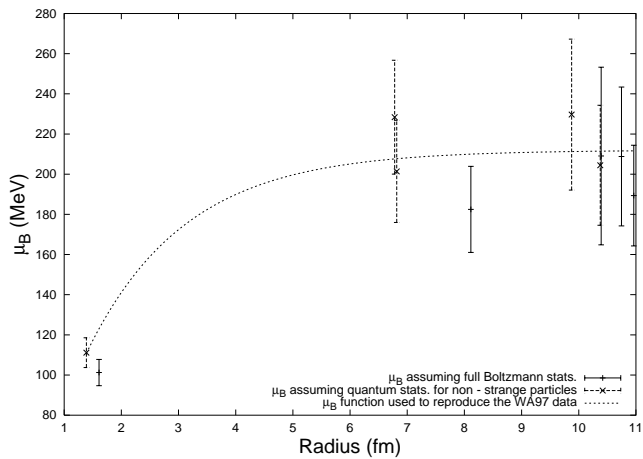}
\caption{\label{fig:mub} Variation of the baryon chemical potential ($\mu_B$) with radius.}
\end{figure}

Figure \ref{fig:mub} shows the variation of the baryon chemical
potential with system size. This function increases rapidly before
saturating. All the Pb+Pb bins are described by the saturation value of
approximately 210 MeV. The obtained values of $\mu_B$ in the
central Pb+Pb bin are in agreement with that of
Becattini et al. \cite{NA49}. The $\mu$ used for the Pb+Pb bins
by Hamieh et al. \cite{hamieh} is just outside one standard deviation of the
values obtained during this analysis. For the p+Be system the
value of $\mu_B$ used in Ref. \cite{hamieh}  (150 MeV) is much larger
than that obtained in this analysis ($111 \pm 7$~MeV) .

The exact variation of $\mu_B$, $T$ and $R$ in the range between the
p--Be and Pb--Pb may differ from what is shown here and new data
from CERN NA57 in this range are eagerly awaited.
The figures mentioned above (\ref{fig:temp}, and \ref{fig:mub}) also show
the variation of the parameters $T$, and $\mu_B$ with centrality in
the case where all particle multiplicities are assumed to be Boltzmann.
These parameters are seen to be in agreement with the parameters
from Analysis A. The value of the $\chi^2$
parameter obtained for the purely Boltzmann fits were of
order 1 for all systems considered.

\subsection{Predicted Evolution of Particle Yields}

\begin{figure*}
\begingroup%
  \makeatletter%
  \newcommand{\GNUPLOTspecial}{%
    \@sanitize\catcode`\%=14\relax\special}%
  \setlength{\unitlength}{0.1bp}%
\begin{picture}(3600,2160)(0,0)%
\includegraphics{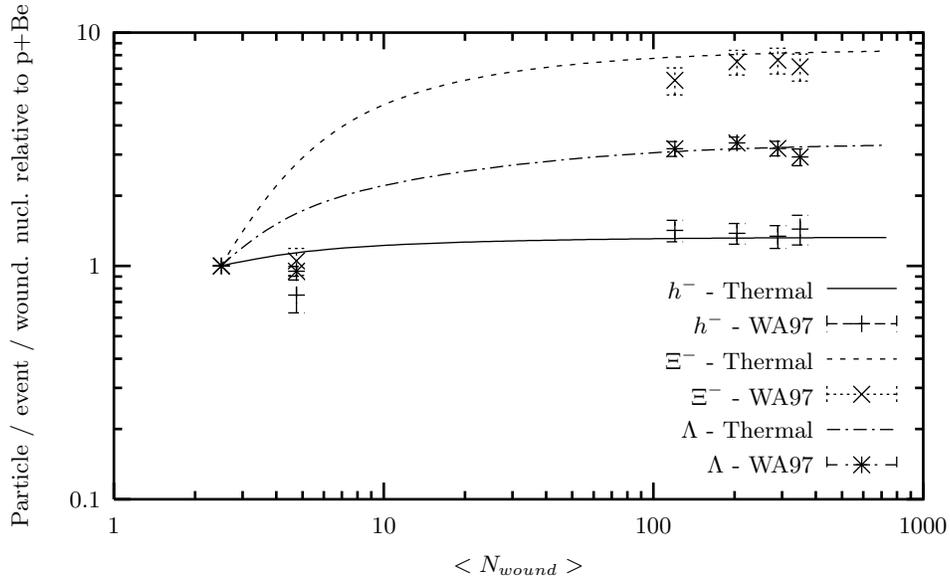}
\put(3037,430){\makebox(0,0)[r]{$\Lambda$ - WA97}}%
\put(3037,563){\makebox(0,0)[r]{$\Lambda$ - Thermal}}%
\put(3037,696){\makebox(0,0)[r]{$\Xi^-$ - WA97}}%
\put(3037,829){\makebox(0,0)[r]{$\Xi^-$ - Thermal}}%
\put(3037,962){\makebox(0,0)[r]{$h^-$ - WA97}}%
\put(3037,1095){\makebox(0,0)[r]{$h^-$ - Thermal}}%
\put(1925,50){\makebox(0,0){$<N_{wound}>$}}%
\put(100,1180){%
\makebox(0,0)[b]{\shortstack{Particle / event / wound. nucl. relative to p+Be}}%
}%
\put(3450,200){\makebox(0,0){1000}}%
\put(2433,200){\makebox(0,0){100}}%
\put(1417,200){\makebox(0,0){10}}%
\put(400,200){\makebox(0,0){1}}%
\put(350,2060){\makebox(0,0)[r]{10}}%
\put(350,1180){\makebox(0,0)[r]{1}}%
\put(350,300){\makebox(0,0)[r]{0.1}}%
\end{picture}%
\endgroup
\caption{\label{fig:WA97f} Comparison of the hadron gas model with exact strangeness conservation  and CERN WA97 data for negatives and strange particles.}
\end{figure*}

\begin{figure*}
\begingroup%
  \makeatletter%
  \newcommand{\GNUPLOTspecial}{%
    \@sanitize\catcode`\%=14\relax\special}%
  \setlength{\unitlength}{0.1bp}%
\begin{picture}(3600,2916)(0,0)%
\includegraphics{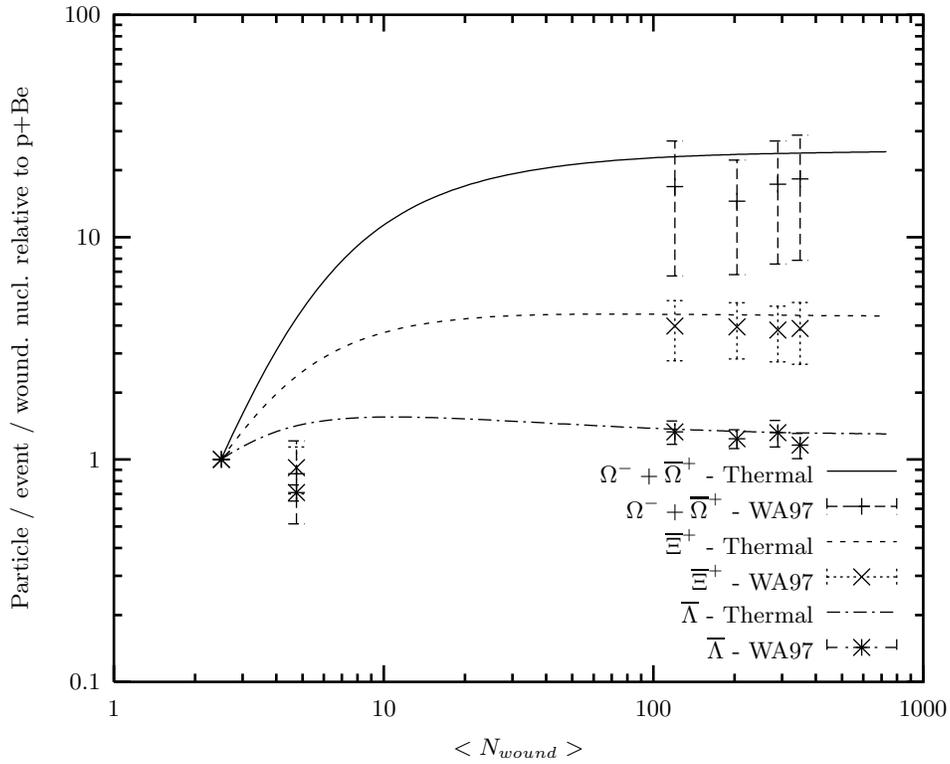}
\put(3037,430){\makebox(0,0)[r]{$\overline{\Lambda}$ - WA97}}%
\put(3037,563){\makebox(0,0)[r]{$\overline{\Lambda}$ - Thermal}}%
\put(3037,696){\makebox(0,0)[r]{$\overline{\Xi}^+$ - WA97}}%
\put(3037,829){\makebox(0,0)[r]{$\overline{\Xi}^+$ - Thermal}}%
\put(3037,962){\makebox(0,0)[r]{$\Omega^- + \overline{\Omega}^+$ - WA97}}%
\put(3037,1095){\makebox(0,0)[r]{$\Omega^- + \overline{\Omega}^+$ - Thermal}}%
\put(1925,50){\makebox(0,0){$<N_{wound}>$}}%
\put(100,1558){%
\makebox(0,0)[b]{\shortstack{Particle / event / wound. nucl. relative to p+Be}}%
}%
\put(3450,200){\makebox(0,0){1000}}%
\put(2433,200){\makebox(0,0){100}}%
\put(1417,200){\makebox(0,0){10}}%
\put(400,200){\makebox(0,0){1}}%
\put(350,2816){\makebox(0,0)[r]{100}}%
\put(350,1977){\makebox(0,0)[r]{10}}%
\put(350,1139){\makebox(0,0)[r]{1}}%
\put(350,300){\makebox(0,0)[r]{0.1}}%
\end{picture}%
\endgroup
\caption{\label{fig:WA97fa}Comparison of the hadron gas model with exact strangeness conservation  and CERN WA97 data for the $\Omega$ and strange anti-particles.}
\end{figure*}

Figs.~\ref{fig:WA97f} and \ref{fig:WA97fa} show the ability of the
model to reproduce the data using the functions in Figs~\ref{fig:temp}
and \ref{fig:mub}, and equation \ref{eq:radius} to predict the
variation in chemical freeze-out temperature $T$, baryon chemical
potential $\mu_B$, and radius $R$ with average number of wounded nucleons.
As is clear, the agreement of the model with experimental data is good.
The exact shape of the thermal model predictions shown
in Figure \ref{fig:WA97f} and \ref{fig:WA97fa} show a large dependence
on the exact relationship between number of wounded nucleons and system
size, and a weaker dependence on the functional fit to the other two
parameters. This is rather unfortunate as the thermal model $R$ values
obtained have uncertainties equal to their magnitude, so this relationship
is not constrained by the model.


\section{Conclusions}
A thermal model conserving baryon number and charge on average, and
strangeness exactly has been used to describe particle yields from heavy ion
collisions. The formalism includes all strange particles. The model has been
applied to the CERN WA97 data, and is shown to be able to
reproduce the data for all centrality
classes (Figs.~\ref{fig:WA97f} and \ref{fig:WA97fa}).
The accuracy of the model parameters (temperature $T$, radius $R$, and baryon
chemical potential $\mu_B$) is restricted by the lack of $4\pi$ integrated
particle yields for multi-strange particles. In order to reproduce WA97
graphical presentation of their data accurately, some knowledge of the
variation of $T$, $R$, and $\mu_B$ was required. A functional form of the
variation of these parameters with the number of wounded nucleons has been
determined and is presented in Figs~\ref{fig:temp} and  \ref{fig:mub}, and
equation \ref{eq:radius}.

The parameters obtained in this model for the central Pb+Pb bins are in
agreement with the $4\pi$ thermal model application of
Becattini et al. \cite{NA49}. This lends some weight to the accuracy of
the model when fitting particle ratios in a limited kinematic region.
This grand canonical model includes the factor $\gamma_S \neq 1$ to predict
the yields of strange particles. To differentiate between canonical
strangeness suppression, and suppression of strange particles by anomalous
phase space occupancy\footnote{parameterised by $\gamma_S \neq 1$ in thermal
models}, the yield of the $\phi$ particle could be used. Yields of
this particle are not sensitive to canonical strangeness suppression, but as
it contains an $s$ and an $\overline{s}$ quark, these yields will be sensitive
to $\gamma_S \neq 1$.

The enhancement of strange particles measured by CERN WA97 has
been considered a signal for QGP formation. At first glance, the
ability of a full equilibrium thermal model to reproduce the data
suggests the existence of a deconfined state, since equilibrium strange
particle yields have been proposed as a possible signal for
deconfinement \cite{rafmul}. A deconfined phase is not, however,
expected to be formed in the p+Be system, as it is only expected
in large dense systems. The ability of the model to reproduce the
p+Be data suggests that it is possible for strange particles to
reach equilibrium yields by hadronic interactions alone.

\begin{acknowledgments}
Sahal Yacoob wishes to thank the National Research Foundation
(NRF) of South Africa for supporting this work.
\end{acknowledgments}

\bibliography{paper3}

\begin{thebibliography}{14}
\expandafter\ifx\csname natexlab\endcsname\relax\def\natexlab#1{#1}\fi
\expandafter\ifx\csname bibnamefont\endcsname\relax
  \def\bibnamefont#1{#1}\fi
\expandafter\ifx\csname bibfnamefont\endcsname\relax
  \def\bibfnamefont#1{#1}\fi
\expandafter\ifx\csname citenamefont\endcsname\relax
  \def\citenamefont#1{#1}\fi
\expandafter\ifx\csname url\endcsname\relax
  \def\url#1{\texttt{#1}}\fi
\expandafter\ifx\csname urlprefix\endcsname\relax\def\urlprefix{URL }\fi
\providecommand{\bibinfo}[2]{#2}
\providecommand{\eprint}[2][]{\url{#2}}

\bibitem[{\citenamefont{Antinori et~al.}(1999)}]{59}
\bibinfo{author}{\bibfnamefont{F.}~\bibnamefont{Antinori}}
  \bibnamefont{et~al.}, \bibinfo{journal}{Nucl.\ Phys.\ A}
  \textbf{\bibinfo{volume}{661}}, \bibinfo{pages}{130c} (\bibinfo{year}{1999}).

\bibitem[{\citenamefont{Hamieh et~al.}(2000)\citenamefont{Hamieh, Redlich, and
  Tounsi}}]{hamieh}
\bibinfo{author}{\bibfnamefont{S.}~\bibnamefont{Hamieh}},
  \bibinfo{author}{\bibfnamefont{K.}~\bibnamefont{Redlich}}, \bibnamefont{and}
  \bibinfo{author}{\bibfnamefont{A.}~\bibnamefont{Tounsi}},
  \bibinfo{journal}{Phys.\ Lett.\ B} \textbf{\bibinfo{volume}{486}},
  \bibinfo{pages}{61} (\bibinfo{year}{2000}).

\bibitem[{\citenamefont{Redlich}(2002)}]{stonybrook}
\bibinfo{author}{\bibfnamefont{K.}~\bibnamefont{Redlich}},
  \bibinfo{journal}{Nucl. Phys.} \textbf{\bibinfo{volume}{A698}},
  \bibinfo{pages}{94} (\bibinfo{year}{2002}).

\bibitem[{\citenamefont{Braun-Munzinger
  et~al.}(2002)\citenamefont{Braun-Munzinger, Cleymans, Oeschler, and
  Redlich}}]{Braun-Munzinger}
\bibinfo{author}{\bibfnamefont{P.}~\bibnamefont{Braun-Munzinger}},
  \bibinfo{author}{\bibfnamefont{J.}~\bibnamefont{Cleymans}},
  \bibinfo{author}{\bibfnamefont{H.}~\bibnamefont{Oeschler}}, \bibnamefont{and}
  \bibinfo{author}{\bibfnamefont{K.}~\bibnamefont{Redlich}},
  \bibinfo{journal}{Nucl.\ Phys.\ A} \textbf{\bibinfo{volume}{697}},
  \bibinfo{pages}{902} (\bibinfo{year}{2002}).

\bibitem[{\citenamefont{Cleymans
  et~al.}(1997{\natexlab{a}})\citenamefont{Cleymans, Redlich, and
  Suhonen}}]{formal}
\bibinfo{author}{\bibfnamefont{J.}~\bibnamefont{Cleymans}},
  \bibinfo{author}{\bibfnamefont{K.}~\bibnamefont{Redlich}}, \bibnamefont{and}
  \bibinfo{author}{\bibfnamefont{E.}~\bibnamefont{Suhonen}},
  \bibinfo{journal}{Z.\ Phys.\ C} \textbf{\bibinfo{volume}{76}},
  \bibinfo{pages}{269} (\bibinfo{year}{1997}{\natexlab{a}}).

\bibitem[{\citenamefont{Cleymans and Redlich}(1999)}]{cleymans-redlich}
\bibinfo{author}{\bibfnamefont{J.}~\bibnamefont{Cleymans}} \bibnamefont{and}
  \bibinfo{author}{\bibfnamefont{.~K.} \bibnamefont{Redlich}},
  \bibinfo{journal}{Phys.\ Rev.\ C} \textbf{\bibinfo{volume}{60}},
  \bibinfo{pages}{054908} (\bibinfo{year}{1999}).

\bibitem[{\citenamefont{Becattini et~al.}(2001)\citenamefont{Becattini,
  Cleymans, Keranen, Redlich, and Suhonen}}]{NA49}
\bibinfo{author}{\bibfnamefont{F.}~\bibnamefont{Becattini}},
  \bibinfo{author}{\bibfnamefont{J.}~\bibnamefont{Cleymans}},
  \bibinfo{author}{\bibfnamefont{A.}~\bibnamefont{Keranen}},
  \bibinfo{author}{\bibfnamefont{K.}~\bibnamefont{Redlich}}, \bibnamefont{and}
  \bibinfo{author}{\bibfnamefont{E.}~\bibnamefont{Suhonen}},
  \bibinfo{journal}{Phys.\ Rev.\ C} \textbf{\bibinfo{volume}{63}},
  \bibinfo{pages}{649} (\bibinfo{year}{2001}).

\bibitem[{\citenamefont{Braun-Munzinger and Stachel}(1996)}]{stachel}
\bibinfo{author}{\bibfnamefont{P.}~\bibnamefont{Braun-Munzinger}}
  \bibnamefont{and} \bibinfo{author}{\bibfnamefont{J.}~\bibnamefont{Stachel}},
  \bibinfo{journal}{Nucl.\ Phys.\ A} \textbf{\bibinfo{volume}{606}},
  \bibinfo{pages}{320} (\bibinfo{year}{1996}).

\bibitem[{\citenamefont{Braun-Munzinger and Stachel}(1999)}]{stachel2}
\bibinfo{author}{\bibfnamefont{P.}~\bibnamefont{Braun-Munzinger}}
  \bibnamefont{and} \bibinfo{author}{\bibfnamefont{J.}~\bibnamefont{Stachel}},
  \bibinfo{journal}{Nucl.\ Phys.\ A} \textbf{\bibinfo{volume}{654}},
  \bibinfo{pages}{119c} (\bibinfo{year}{1999}).

\bibitem[{\citenamefont{Sollfrank et~al.}(1994)}]{S+S}
\bibinfo{author}{\bibfnamefont{J.}~\bibnamefont{Sollfrank}}
  \bibnamefont{et~al.}, \bibinfo{journal}{Z.\ Phys.\ C}
  \textbf{\bibinfo{volume}{61}}, \bibinfo{pages}{659} (\bibinfo{year}{1994}).

\bibitem[{\citenamefont{Barton et~al.}(2001)}]{barton}
\bibinfo{author}{\bibfnamefont{R.}~\bibnamefont{Barton}} \bibnamefont{et~al.},
  \bibinfo{journal}{J.\ Phys.\ G} \textbf{\bibinfo{volume}{27}},
  \bibinfo{pages}{367} (\bibinfo{year}{2001}).

\bibitem[{\citenamefont{Cleymans
  et~al.}(1997{\natexlab{b}})\citenamefont{Cleymans, Keranen, Marais, and
  Suhonen}}]{marais}
\bibinfo{author}{\bibfnamefont{J.}~\bibnamefont{Cleymans}},
  \bibinfo{author}{\bibfnamefont{A.}~\bibnamefont{Keranen}},
  \bibinfo{author}{\bibfnamefont{M.}~\bibnamefont{Marais}}, \bibnamefont{and}
  \bibinfo{author}{\bibfnamefont{E.}~\bibnamefont{Suhonen}},
  \bibinfo{journal}{Phys.\ Rev.\ C} \textbf{\bibinfo{volume}{56}},
  \bibinfo{pages}{2747} (\bibinfo{year}{1997}{\natexlab{b}}).

\bibitem[{\citenamefont{Rafelski and Muller}(1982)}]{rafmul}
\bibinfo{author}{\bibfnamefont{J.}~\bibnamefont{Rafelski}} \bibnamefont{and}
  \bibinfo{author}{\bibfnamefont{B.}~\bibnamefont{Muller}},
  \bibinfo{journal}{Phys.\ Rev.\ Lett.} \textbf{\bibinfo{volume}{48}},
  \bibinfo{pages}{1066} (\bibinfo{year}{1982}).

\bibitem[{\citenamefont{Yen and Gorenstein}(1999)}]{gorenstein2}
\bibinfo{author}{\bibfnamefont{G.~D.} \bibnamefont{Yen}} \bibnamefont{and}
  \bibinfo{author}{\bibfnamefont{M.~I.} \bibnamefont{Gorenstein}},
  \bibinfo{journal}{Phys.\ Rev.\ C} \textbf{\bibinfo{volume}{59}},
  \bibinfo{pages}{2788} (\bibinfo{year}{1999}).

\end{thebibliography}
\end{document}